\documentclass{achemso}
\usepackage[utf8]{inputenc}
\usepackage[T1]{fontenc}
\usepackage{amsmath}
\usepackage{siunitx}

\usepackage{xcolor}  
\usepackage{graphicx}

\graphicspath{{Images/}}

\author{David Rosser}
\affiliation{Department of Physics, University of Washington, Seattle, Washington 98195, USA}
\author{Dario Gerace}
\affiliation{Dipartimento di Fisica, Universit\`{a} di Pavia, via Bassi 6, 27100 Pavia, Italy }
\author{Lucio C. Andreani}
\affiliation{Dipartimento di Fisica, Universit\`{a} di Pavia, via Bassi 6, 27100 Pavia, Italy }
\author{Arka Majumdar}
\affiliation{Department of Physics, University of Washington, Seattle, Washington 98195, USA}
\affiliation{Department of Electrical and Computer Engineering, University of Washington, Seattle, Washington 98195, USA}
\email{arka@uw.edu}

\title{\textbf{Optimal condition to probe strong coupling of two-dimensional excitons and zero-dimensional cavity modes }}

\begin{document}

\begin{abstract}
The light-matter interaction associated with a two-dimensional (2D) excitonic transition coupled to a zero-dimensional (0D) photonic cavity is fundamentally different from coupling localized excitations in quantum dots or color centers, which have negligible spatial extent compared to the cavity-confined mode profile. By calculating the radiation-matter coupling of the exciton transition of a surface deposited 2D material and a 0D photonic crystal nanobeam mode, we found that there is an optimal spatial extent of the monolayer material that maximizes such an interaction strength due to the competition between minimizing the excitonic envelope function area and maximizing the total integrated field. This is counter to the intuition from the Dicke model, where the oscillator strength is expected to monotonically grow with the number of oscillators, which correlates to the monolayer area assuming the excitonic wavefunction is delocalized over the entire quantum well. We also found that at near zero exciton-cavity detuning, the direct transmission efficiency of a waveguide-integrated cavity can be severely suppressed, which suggests performing experiments by using a side-coupled cavity to get better performances.
\end{abstract}

\maketitle

\section{Introduction}
Realizing single-photon nonlinear optics in a scalable platform could revolutionize both classical and quantum information science and engineering\cite{miller_attojoule_2017, sun_single-photon_2018, englund_ultrafast_2012, chang_single-photon_2007}. Among various systems being explored to reach this strongly nonlinear regime, cavity integrated excitonic materials show great promise. Quantum confinement of the exciton wave function provides an enhanced density of states, which allows strong radiation-matter interaction. This interaction can be further enhanced by integrating the material into a wavelength-scale photonic cavity for temporal and spatial confinement of the electromagnetic field. In general, the radiation-matter coupling depends on the dimensionality of both exciton and photon fields, and it has been accurately predicted \cite{andreani_exciton-polaritons_2013}. Furthermore, nonlinear interactions, derived from Coulomb contributions among the charged particles, are enhanced due to the quantum confinement \cite{carusotto_quantum_2013}, which holds promise for application of these systems as analog quantum simulators \cite{angelakis_book_2017}.

To reach this advantageous nonlinear regime, the cavity and the exciton must be strongly coupled, i.e., the coherent coupling strength between the two oscillators should be larger than the system losses. In this regime, the cavity-confined photons and the excitons are hybridized to create a new elementary excitation, known as polariton, whose properties crucially depend on the dimensionality of the exciton and photon degrees of freedom. Strong coupling and subsequent single photon nonlinear optics have been demonstrated in self-assembled quantum dots coupled to zero-dimensional (0D) cavity systems \cite{faraon_coherent_2008, majumdar_probing_2012, reinhard_strongly_2012, najer_gated_2019}. In a quantum dot, the exciton is confined in all three dimensions, which is often defined a 0D exciton. Similarly, in a photonic crystal defect cavity \cite{Notomi_2010} or a fiber-distributed Bragg reflector (DBR) cavity \cite{Hunger_2010}, light is confined in wavelength scale in all three dimensions, making these systems 0D cavities. While such 0D polaritons can provide the strongest nonlinearity, arising from the quantum anharmonicity induced by the 0 exciton \cite{Hennessy_Nature_2007,najer_gated_2019}, practical limitations, such as dispersionless cavities and the stochastic nature of the quantum dots, prevent the scalability of such a platform. Another well-studied polaritonic system consists of two-dimensional (2D) excitons in quantum wells integrated with  2D cavities, such as a DBR cavity or nonlocal metasurfaces \cite{byrnes_excitonpolariton_2014, liu_strong_2015, chen_metasurface_2020}. While many quantum optical effects have been predicted in these 2D exciton-polariton systems \cite{Gerace_nphys_2009,Liew_coupled_modes_2010}, the lack of excitonic wave function confinement in all three dimensions precluded a clear observation of single photon nonlinearity, i.e. reaching the regime of polariton blockade under resonant excitation \cite{verger_polariton_2006}.

Recently, signatures of single photon nonlinearity have been reported in a III-V quantum well system coupled to an optically confined mode of curved fiber-DBR \cite{delteil_towards_2019, munoz-matutano_emergence_2019}. While these works provide remarkable proof-of concept demonstrations with promising perspectives \cite{Gerace_NV_2019}, the in-situ tuning advantage of a fiber-DBR cavity comes at the expense of a larger mode volume as compared to a photonic crystal defect cavity \cite{greuter_small_2014, fryett_encapsulated_2018}, as well as an unclear path for scaling to a cavity array. As such, on-chip 0D sub-wavelength mode volume cavities coupled to a 2D excitonic transition can simultaneously provide a large light-matter interaction and a clear path to a scalable architecture. However, strong coupling between such an on-chip 0D cavity mode and 2D exciton has not been demonstrated, yet. A primary difficulty to achieve such an accomplishment has proven to be the inevitable deterioration of quantum well excitons due to etching, when inorganic semiconductor material platforms are used. This problem can be alleviated by using atomically thin van der Waals materials, such as transition metal dichalcogenides (TMD), as they can be transferred on a pre-fabricated photonic crystal cavity. However, even though these materials have long been integrated with 0D on-chip photonic crystal defect cavities \cite{Dirk_PhC_MoS2,Wu_2014}, till date there has been no report on radiation-matter strong coupling. 

In this work, we theoretically analyze such a system to elucidate the conditions allowing to experimentally probe the strong coupling regime between 2D excitons and the 0D cavity mode of a sub-wavelength photonic nanocavity. More specifically, we considered the neutral exciton in a monolayer TMD, such as $\mathrm{MoSe}_{2}$, deposited on a $\mathrm{SiN}$ photonic crystal nanobeam cavity as a model system to be quantitatively analyzed. We have numerically calculated the radiation-matter coupling depending on the specific near field profile of the confined mode, and we show that there is an optimal spatial coverage of the 2D exciton in such 0D cavity when the light-matter interaction is maximized. Additionally, applying an input-output approach to calculate the cavity transmission, we show that due to the absorption from 2D material, the cavity transmission drops significantly for a resonant exciton-photonic crystal cavity system. By exploiting side coupling between a waveguide and the cavity containing the excitonic material, strong coupling could be probed in a transmission configuration. 

\section{Theoretical model}
A 2D exciton coupled to a 0D cavity mode in the weak excitation regime can be modeled with the phenomenological Hamiltonian describing two coupled oscillators (in a frame rotating at the frequency of an external pump laser) \cite{gerace_quantum_2007, deng_exciton-polariton_2010}

\begin{equation}\label{ham}
	H_{XC} = \hbar \Delta_{XL} \hat{a}^{\dagger} \hat{a} + \hbar \Delta_{CL} \hat{c}^{\dagger} \hat{c} + \hbar g (\hat{a}^{\dagger} \hat{c} + \hat{c}^{\dagger} \hat{a}).
\end{equation}
with a coherent exciton-cavity coupling rate $g$, where $\Delta_{XL} = \omega_{X} - \omega_{L}$ and $\Delta_{CL} = \omega_{C} - \omega_{L}$ are the detunings of the excitonic transition $(\omega_X)$ and cavity mode $(\omega_C)$ from the laser frequency ($\omega_{L}$), respectively; $\hat{a}$ ($\hat{c}$) is the bosonic annihilation operator for the exciton (cavity) mode. We are only concerned with the exciton mode with the same spatial wave function as the cavity mode due to the limited dispersion of a 0D cavity \cite{verger_polariton_2006}. Hence, we neglect the in-plane momentum distribution of the 2D exciton in our model.

\begin{figure}
	\centering
	\includegraphics[width=1.0\linewidth]{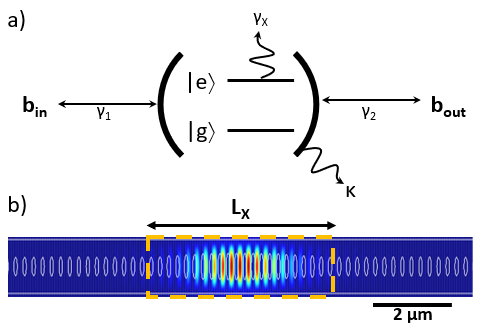}
	\caption{a) The input-output schematic of an in-line cavity coupled to an optical transition. $\kappa$ and $\gamma_{X}$ are the intrinsic losses of the cavity and exciton, respectively. And $\gamma_{1,2}$ are the waveguide-coupled losses. b) Electric field intensity simulated at the center of a SiN nanobeam cavity by a FDTD electromagnetic solver at the cavity mode resonant frequency, showing wavelength scale field confinement. The maximum field intensity is seen in the center of the nanobeam. $L_{X}$ is the length of the integrated quantum well.}
	\label{fig:003figure1c}
\end{figure}

We estimate the light-matter interaction strength $g$ between the 2D exciton and the 0D cavity by noting that the dielectric function of monolayer $\mathrm{MoSe}_{2}$ can be modeled as a Lorentzian oscillator $\varepsilon (\omega) = \varepsilon_{b} + \frac{A}{\omega_{X}^{2} - \omega^{2} - i \gamma_{X} \omega}$ \cite{lundt_monolayered_2016}, where $\varepsilon_{b} = 26$ is the background dielectric constant of the TMD layer \cite{morozov_optical_2015}, which results in a perturbative shift of the cavity resonance; $A$ is an ``effective'' oscillator strength having dimensions of a squared energy, $\omega_{X}$ is the energy of the excitonic transition (with $\hbar=1$), and $\gamma_{X}$ is the total exciton loss rate (including both radiative and non-radiative contributions). In the following, we fix $A = 0.4 \; \si{\electronvolt^{2}}$ as a representative value from experimental reflectivity measurements \cite{lundt_monolayered_2016}. By treating the monolayer TMD as a delocalized semiconductor quantum well exciton in a dielectric medium coupled to a confined optical mode of the cavity a concise expression for the light-matter coupling energy can be derived (App.~A).

\begin{equation}\label{rmc}
	g = \frac{\sqrt{A}}{2} \sqrt{\frac{L_{z}}{L_{\mathrm{eff}}}}.
\end{equation}
$A$ is the effective oscillator strength contained in the dielectric constant, $L_{z}$ is the thickness of the monolayer material, and $L_{\mathrm{eff}}$ is a length scale defined by the competition between minimizing the excitonic envelope function area and maximizing the total integrated field.

\begin{equation}\label{ECC}
	L_{\mathrm{eff}} = \left ( \frac{1}{L_{x} L_{y}} \int_{0}^{L_{x}} dx \int_{0}^{L_{y}} dy \left[ \mathbf{E}_{norm,x}^{(2D)} (x,y) + \mathbf{E}_{norm,y}^{(2D)} (x,y) \right] \right )^{-2}
\end{equation}
$L_{x}$ ($L_{y}$) is the length (width) of the integrated monolayer material (Fig. \ref{fig:003figure1c}b) and $\mathbf{E}_{norm,x}^{(2D)}$ ($\mathbf{E}_{norm,y}^{(2D)}$) is the normalized electromagnetic field in the $x$ ($y$) direction.

\begin{figure}
	\centering
	\includegraphics[width=1.0\linewidth]{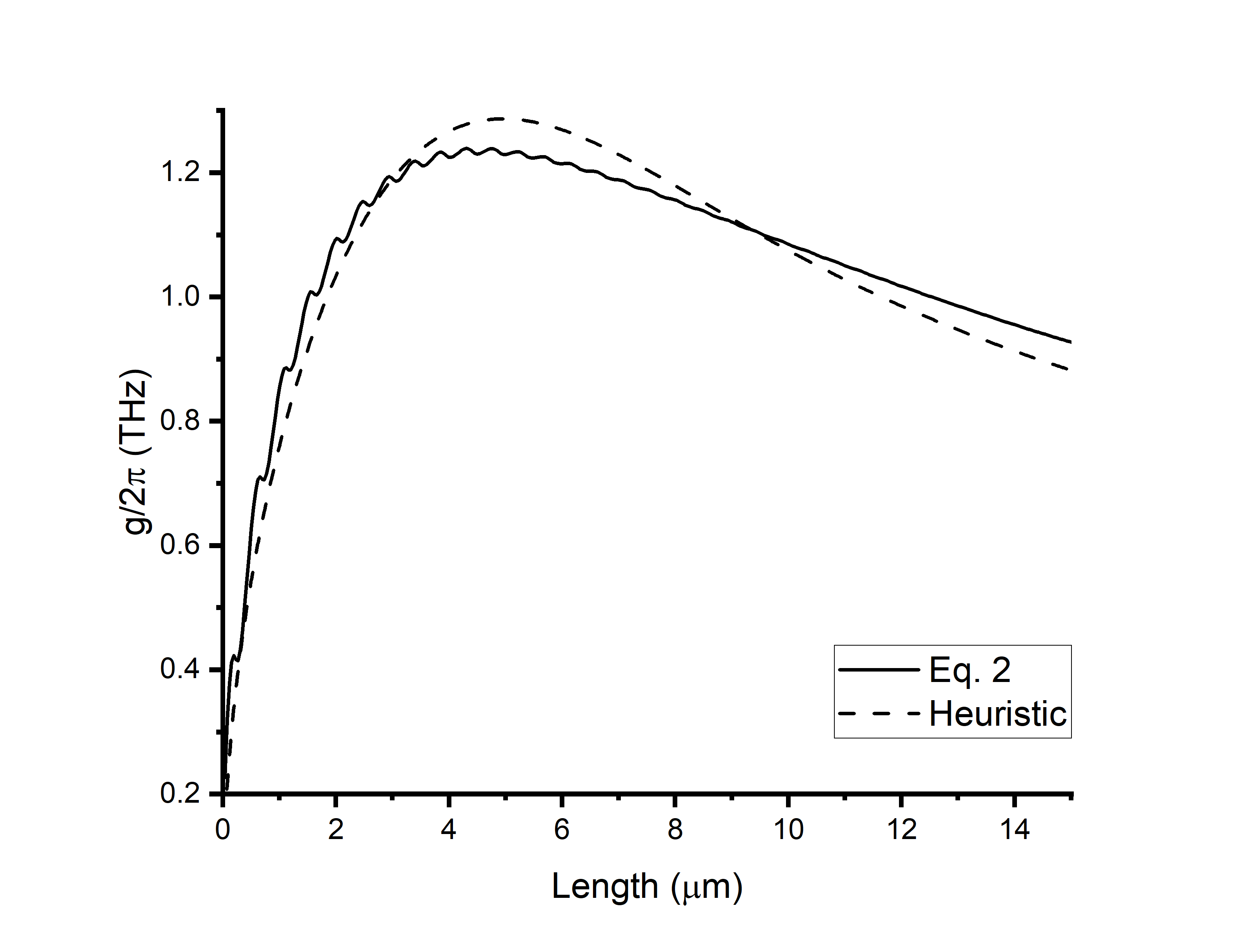}
	\caption{Light-matter coupling for different lengths $L_{X}$ of the integrated quantum well with $L_{Y}$ fixed to the width of the waveguide. The oscillations seen in the radiation matter coupling originate from the periodic variation of the electric field commensurate with the lattice spacing of the nanobeam air holes. $g/2\pi = 1.2389 \; \si{\tera \hertz}$ is the maximum value for this cavity design and oscillator strength. The dotted line is a fit to the heuristic equation in the main text elucidating the peak in the light-matter coupling for a cavity confinement length of $\sigma=1.77$ \si{\micro \meter}.}
	\label{fig:gvsl}
\end{figure}

While this formalism can be applied to any extended 2D coherent media in confined cavity geometries, we illustrate this result assuming parameters appropriate for a $\mathrm{MoSe}_{2}$ monolayer deposited on a SiN nanobeam cavity, because such a system can be readily fabricated in practice \cite{rosser_dispersive_2020}. Using a finite difference time domain (FDTD) electromagnetic solver (from Lumerical-Ansys), we calculate the cavity field profile (Fig. \ref{fig:003figure1c}b) to be used into Eq. \ref{ECC} with a resonance at $\omega_{C}/2 \pi = 395.777 \; \si{\tera \hertz}$ (wavelength of  $757 \; \si{\nano \meter}$) (Appendix B). Taking the effective thickness of the monolayer material to be equal to the measured one, $t_{\mathrm{MoSe}_{2}} = 0.7$ \si{\nano \meter}, we find a maximal value for the light-matter coupling with monolayer length of $4.31$ \si{\micro \meter} (Fig. \ref{fig:gvsl}). This result runs counter to the intuition from the Dicke model, in which a giant oscillator is expected to grow monotonically with the number of oscillators $(g \propto \sqrt{N} g_{0})$ \cite{garraway_dicke_2011}, which in this case, correlates to the area of monolayer $\mathrm{MoSe}_{2}$ assuming the excitonic wavefunction is delocalized over the entire field integration region.

Heuristically, we can understand this optimal overlap between the 2D exciton envelope function and the cavity field profile in terms of the light-matter coupling by recognizing that steady state electric field of the nanobeam cavity has an approximately Gaussian envelope along the cavity axis, with a width $\sigma$ (units of length) modulated by a sinusoidal signal of the photonic lattice periodicity \cite{quan_deterministic_2011} (see, e.g., Fig. \ref{fig:003figure1c}a). Assuming the length of the coherent polarization due to the delocalized excitonic wavefunction is same as the length of the cavity integration, $L_{x}$, substitution of a Gaussian cavity field profile into Eq. \ref{ECC} gives a light-matter interaction strength of the form $g \propto \frac{1}{\sqrt{L_{x}}} \int_{-L_{x}/2}^{L_{x}/2} e^{-\frac{1}{2} \left( \frac{x}{\sigma} \right)^{2}} dx \propto \frac{\sigma}{\sqrt{L_{x}}} \mathrm{erf} (\frac{L_{x}}{2 \sqrt{2} \sigma})$. The latter function gives a peak in the light-matter coupling around $2.80 \sigma$, which roughly corresponds with the $2.44 \sigma$ that is numerically calculated for the designed nanobeam cavity. In Fig.\ref{fig:gvsl}, we overlay this estimate on top of the numerical simulation, for the sake of clarity and completeness.

We now discuss the experimental scheme allowing to probe these excitations in the system. Often, such radiation-matter coupled systems are measured via incoherent photoluminescence; however, coherent driving in the transmission configuration is necessary in view of practical development of quantum technology applications \cite{englund_resonant_2010}. For the on-chip configuration, the exciton-polariton modes are generally probed using a two-sided cavity \cite{collett_squeezing_1984} (Fig. \ref{fig:003figure1c}a). An input grating is used to send light to the coupled system and the transmitted light is collected via an output grating. Using the input-output formalism \cite{gardiner_input_1985} a simple transmission function for the system described by the Hamiltonian in Eq. \ref{ham}, can be derived (Appendix C):
\begin{equation}\label{trans}
	T(\omega) = \frac{\gamma_{1} \gamma_{2}}{ \left[ \omega - \omega_{C} - \frac{ (\omega - \omega_{X}) g^2 }{ (\omega - \omega_{X})^2 + \gamma_{X}^2} \right ]^2+\left[ \frac{1}{2}( \gamma_{1} + \gamma_{2} ) + \kappa + \frac{ \gamma_{X} g^2 }{ (\omega - \omega_{X})^2 + \gamma_{X}^2} \right ]^2}.
\end{equation}
where, we include intrinsic cavity losses $\kappa$, cavity coupling to the input (output) waveguide $\gamma_1 (\gamma_2)$, and excitonic losses $\gamma_{X}$. In our model system, the nanobeam cavity is symmetrically coupled to the waveguide (i.e., $\gamma_1=\gamma_2=\gamma$).

The intrinsic cavity loss and cavity-waveguide coupling can be inferred from the FDTD simulations. The designed nanobeam cavity has a loaded quality factor of $Q_{loaded}=11924$ and an intrinsic quality factor of $Q_{intrinsic}=25480$. The intrinsic quality factor of the cavity is found by increasing the number of Bragg mirror holes until the waveguide is no longer coupled to the cavity and the simulated quality factor approaches an asymptotic value. We note that for this particular cavity, we are choosing an on-substrate SiN cavity due to its mechanical stability \cite{fryett_encapsulated_2018}, hence the reduced quality factor compared to a suspended nanobeam cavity. The decay rate of the cavity field is $\kappa = \frac{1}{2} \frac{\omega_{C}}{Q_{intrinsic}} = 2 \pi \times 7.77$ \si{\giga \hertz}. Similarly, the decay rate of the loaded cavity field is $\kappa + \gamma = \frac{1}{2} \frac{\omega_{C}}{Q_{loaded}} = 2 \pi \times 16.6$ \si{\giga \hertz}, which gives a waveguide-coupled field decay rate $\gamma = 2 \pi \times 8.83$ \si{\giga \hertz}. This results in an estimated maximum transmission efficiency of $T_{max} = \left ( \frac{\gamma}{\kappa + \gamma} \right )^{2} = 0.28$ \cite{xu_scattering-theory_2000}.

To probe the system in the strong coupling regime we need to calculate the transmission efficiency for a resonant exciton-cavity system. The temperature-dependent excitonic loss can be approximated using the Rudin equation $\gamma_{X} (T) = \frac{1}{2} [ \gamma_{0} + c_{1} T + \frac{c_{2}}{e^{\Omega/{k_{B} T}} - 1}]$ where $\gamma_{0}$ is the intrinsic linewidth, $c_{1}$ includes exciton interactions with acoustic phonons, $c_{2}$ includes exciton interactions with longitudinal-optical phonons, and $\Omega$ is the average phonon energy \cite{rudin_temperature-dependent_1990}. We fix $\hbar \gamma_{0} = 4.3$ \si{\milli \electronvolt}, $c_{1} = 91$ \si{\micro \electronvolt \kelvin^{-1}}, $c_{2} = 15.6$ \si{\milli \electronvolt}, and $\Omega = 30$ \si{\milli \electronvolt} as representative values for unencapsulated monolayer $\mathrm{MoSe}_{2}$ \cite{selig_excitonic_2016}. We use $\gamma_{X}$ at 4.2 \si{\kelvin} is $2 \pi \times 566$ \si{\giga \hertz} as a representative value for the excitonic linewidth in the strong coupling regime. 

\begin{figure}
	\centering
	\includegraphics[width=1.0\linewidth]{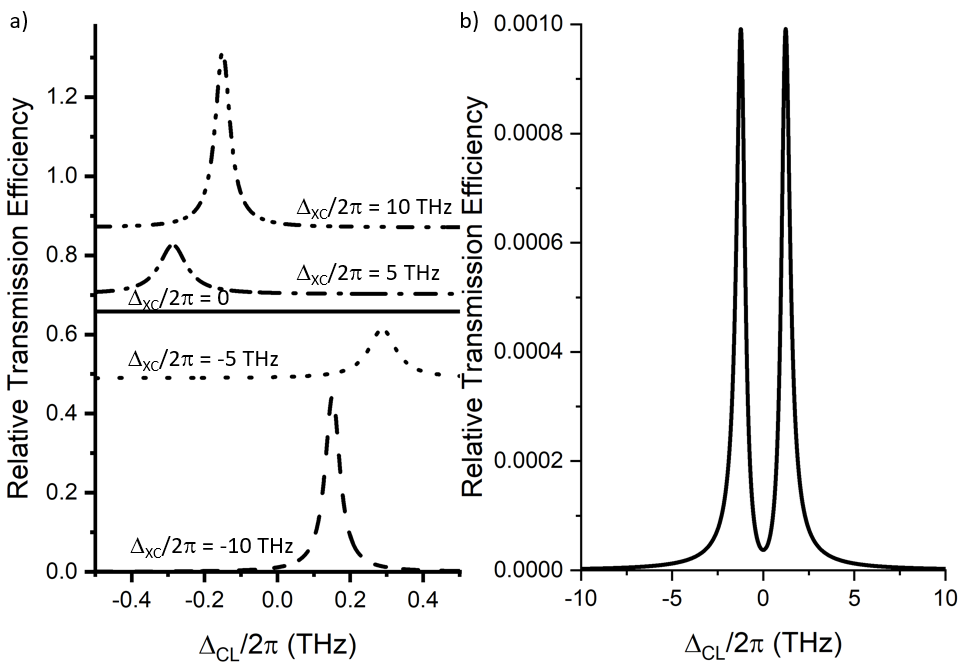}
	\caption{a) Transmission spectrum relative to $T_{max}$ ($T(\omega) / T_{max}$) at different exciton-cavity detunings $\Delta_{XC}$. $\Delta_{CL} = \omega_{C} - \omega_{L}$ is the laser detuning from the bare cavity resonance. b) Transmission spectrum relative to $T_{max}$ at zero exciton-cavity detuning. The solid line in part (b) is the magnified solid line of part (a). Parameters: $\kappa/2\pi = 7.77 \; \si{\giga \hertz}$, $\gamma/2\pi = 8.83 \; \si{\giga \hertz}$, $\gamma_{X}/2\pi = 566 \; \si{\giga \hertz}$, $g/2\pi = 1.2389 \; \si{\tera \hertz}$.}
	\label{fig:003figure3}
\end{figure}

With these values we calculate the transmission spectrum of the coupled exciton-cavity system (Fig. \ref{fig:003figure3}a). We find that, at large exciton-cavity detuning, the transmission efficiency approaches the bare cavity value $T_{max}$.  At smaller detunings, the dispersive cavity shift is noticeable with broadening of the transmission peak. Near zero detuning, however, the intensity of the transmission peak is reduced by several orders of magnitude than the bare cavity transmission in the strong coupling regime (Fig. \ref{fig:003figure3}b). Substituting the parameters, for example, from Fig. \ref{fig:003figure3} into Eq. \ref{Tmaxg} we find the maximum transmission efficiency with the integrated 2D exciton relative to the bare cavity transmission maximum is only $0.098\%$. Thus a major drawback of an in-line symmetric two-sided cavity is the drastic suppression of transmission near zero exciton-cavity detuning. Note that, we experimentally observed a similar reduction in cavity transmission in our exciton-cavity system\cite{rosser_dispersive_2020}. 

\begin{figure}
	\centering
	\includegraphics[width=0.7\linewidth]{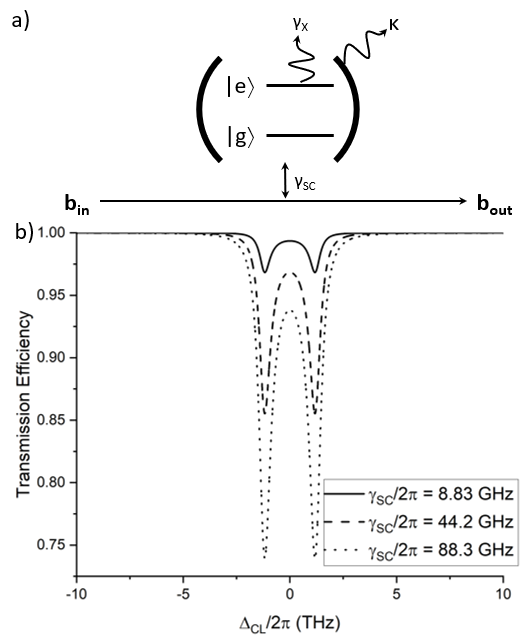}
	\caption{a) The input-output schematic of a side-coupled cavity with an integrated optical transition. $\kappa$ and $\gamma_{X}$ are the intrinsic losses of the cavity and exciton, respectively. And $\gamma_{SC}$ is the waveguide-coupled loss for the side-coupled cavity. b) Side-coupled transmission spectrum for increasing values of the waveguide-coupled loss. Parameters: $\kappa/2\pi = 7.77 \; \si{\giga \hertz}$, $\gamma_{X}/2\pi = 566 \; \si{\giga \hertz}$, $g / 2\pi = 1.2389 \; \si{\tera \hertz}$.}
	\label{fig:003figure5}
\end{figure}

The difficulty of demonstrating high transmission efficiency of the cavity mode near zero exciton-cavity detuning stems from the lack of impedance matching between the loss of the hybridized polariton mode and waveguide-coupled loss. As long as the exciton loss is significantly greater than the cavity loss $\kappa + \gamma$, then we can modify the waveguide-coupled loss to our advantage in measuring the transmission spectrum. This can be achieved by reducing the number of Bragg mirror holes possibly on one or both sides of the nanobeam cavity. Generically, this will lead to a reduction of the cavity quality factor. Alternatively, we can consider a side-coupled nanobeam cavity \cite{afzal_efficient_2019} or ring resonator \cite{rosser_high-precision_2020}. By modifying the width and gap of the coupled waveguide to the nanobeam cavity the waveguide-coupled loss can also be increased for the observation of strong coupling in a waveguide-integrated platform (Fig. \ref{fig:003figure5}). The side-coupled geometry decouples the intrinsic cavity quality factor and field profile from the transmission properties of the cavity design.

\section{Discussion}
We have estimated an optimal length of monolayer $\mathrm{MoSe}_{2}$ for the integration onto a nanobeam cavity. The successful study of polariton physics in this material platform will likely require ex-situ etching due to the size of the monolayer and positional accuracy. Despite the improved cooperativity ($C=\frac{g^2}{\gamma_{X} (\kappa + \gamma)}$) found by maximizing the light-matter interaction, the small transmission efficiency remains a challenge to experimentally probe the strong coupling regime \cite{rosser_high-precision_2020}. This low transmission efficiency may be avoided by decoupling the waveguide-coupled loss from the intrinsic cavity loss by using a side-coupled nanobeam or ring resonator \cite{xu_scattering-theory_2000, afzal_efficient_2019}. This allows for an extra degree of freedom to increase the waveguide-coupled loss at a similar rate to that of the cavity broadening from the perturbing monolayer $\mathrm{MoSe}_{2}$. The limiting factor in this system is the linewidth of the neutral exciton in monolayer $\mathrm{MoSe}_{2}$. hBN encapsulation is a means to narrow the linewidth by modifying the dielectric environment and reducing sample inhomogeneity \cite{martin_encapsulation_2020}. However, experiment may be better served by pursuing two-dimensional excitonic transitions with intrinsically narrow linewidths \cite{kang_coherent_2020}.

\section*{Acknowledgments}
The research was supported by NSF-1845009 and NSF-ECCS-1708579. D.R. is partially supported by a CEI graduate fellowship.

\section*{Appendix A: Light-Matter Coupling}
\setcounter{equation}{0}
\renewcommand{\theequation}{A\arabic{equation}}

The light-matter coupling in the dipole approximation for an excitonic transition in a quantum well (QW) can be written \cite{andreani_strong-coupling_1999, panzarini_quantum_1999, gerace_quantum_2007}
\begin{equation}\label{rmc_begin}
	\hbar g = \left(\frac{\hbar^2 e^2}{4 \varepsilon_{0} m_{0}} f_{xy}\right)^{1/2} \int_{\Sigma} \mathrm{d}x \mathrm{d}y \; \hat{\mathbf{d}} \cdot \mathbf{E}_{norm}^{2D}(x,y)F_{exc}(x,y)
\end{equation}
where $\hbar$ is the reduced Planck constant,  $e$ is the charge of an electron, $\epsilon_o$ is the vacuum permittivity, $f_{xy}$ is an oscillator strength per unit area, $m_{0}$ is the free electron mass, $\hat{\mathbf{d}}$ is a unit vector pointing primarily in the plane of the quantum well, $\mathbf{E}_{norm}^{2D}$ is the normalized electric field at the surface of the nanobeam cavity, and $F_{exc}$ is the normalized exciton envelope function. The integration is performed over the whole cavity region.

We choose to normalize the electric field such that $\mathbf{E}_{norm}^{2D}(x,y) = \frac{1}{\mathcal{\sqrt{N}}}\mathbf{E}(x,y,z_{QW})$ and $\mathcal{N} = \int \varepsilon (\mathbf{r}) \left| \mathbf{E} (\mathbf{r}) \right|^2$. $z_{QW}$ is the z-coordinate of the 2D exciton and $\mathbf{E}(x,y,z)$ is the electric field calculated by a FDTD electromagnetic solver at the cavity mode resonant frequency. The normalized exciton envelope function is defined as $F_{exc} = 1/\sqrt{S}$ where $S$ is the effective area of the excitonic transition, which we take to be the physical area of the monolayer material $S = L_{X} L_{Y}$, assuming the excitonic wavefunction is delocalized over the whole monolayer area. These definitions lead to the effective length scale of Eq. \ref{ECC}.

An equivalent expression for the dielectric function of a quantum well is \cite{andreani_exciton-polaritons_2013}

\begin{align}
	\varepsilon (\omega) & = \varepsilon_{b} \left( 1 + \frac{E_{L}^2 - E_{X}^2}{E_{X}^2 - (\hbar \omega)^2 - i \hbar^2 \gamma_{X} \omega} \right) \\
	& \simeq \varepsilon_{b} \left( 1 + \frac{2 E_{LT} E_{X}}{E_{X}^2 - (\hbar \omega)^2 - i \hbar^2 \gamma_{X} \omega} \right)
\end{align}
where $E_{LT} = \hbar^2 e^2 / (2 \varepsilon_{0} \varepsilon_{b} m_{0} E_{X} L_{z})$ is the longitudinal-transverse splitting. $L_{z}$ is an effective thickness that accounts for the finite penetration of the exciton envelope function into the barriers of the quantum well. The oscillator strength per unit area can then be determined from the oscillator strength measured from reflectivity measurements ($f_{xy} = \frac{\varepsilon_{0} m_{0} L_{z}}{\hbar^2 e^2} A$, where $A$ is defined in the main text as the effective oscillator strength in the Lorentz oscillator expression) \cite{lundt_monolayered_2016}. The compact expression for the light-matter coupling in Eq. \ref{rmc} follows from substitution of this result into Eq. \ref{rmc_begin} with the definition of $L_{eff}$.

\section*{Appendix B: Cavity Example}
\setcounter{equation}{0}
\renewcommand{\theequation}{B\arabic{equation}}

A photonic crystal nanobeam cavity is chosen for its large quality factor, small mode volume and high on-resonance transmission efficiency \cite{quan_deterministic_2011}. We emphasize that the formalism presented in this paper can be used for any other cavities. However, to calculate the light-matter interaction strength we need to use the cavity field profile of a specific cavity design. The one-dimensional photonic crystal defect cavity, also known as a nanobeam cavity, is made of a $t_{\mathrm{waveguide}} = 220$ \si{\nano\meter} thick and  $w_{\mathrm{waveguide}} = 779$ \si{\nano\meter} wide silicon nitride film on silicon oxide substrate. From the center of the nanobeam, where the light is confined, there are 10 tapering holes and 20 Bragg mirror holes. All of the holes are elliptical with a minor axis radius fixed to $40$ \si{\nano\meter}. The tapering holes begin with a $178$ \si{\nano\meter} major axis diameter and a $215$ \si{\nano\meter} center-to-center distance. The tapering region is quadratically tapered to a $121$ \si{\nano\meter} major axis radius and a $233$ \si{\nano\meter} center-to-center distance. The Bragg  mirror region has a major axis radius fixed to $121$ \si{\nano\meter} and a $233$ \si{\nano\meter} center-to-center distance. The performance of the nanobeam cavity was optimized using Lumerical's finite-difference time-domain (FDTD) electromagnetic solver. The dimensions are identical to the cavity found in an experimental dispersive coupling result \cite{rosser_dispersive_2020}. 

\section*{Appendix C: Input-Output Formalism}
\setcounter{equation}{0}
\renewcommand{\theequation}{C\arabic{equation}}

The input $b_{in}$ and output $b_{out}$ fields obey the Heisenberg equations

\begin{align}
	\hat{b}_{out} & = \sqrt{\gamma_{2}} \hat{c} \\
	\frac{d\hat{c}}{dt} & = -\frac{i}{\hbar} [\hat{c},H_{XC}] - \frac{\gamma_{1}}{2} \hat{c} - \frac{\gamma_{2}}{2} \hat{c} + \sqrt{\gamma_{1}} \hat{b}_{in} \\
	\frac{d\hat{a}}{dt} & = -\frac{i}{\hbar} [\hat{a},H_{XC}]
\end{align}
The transmission spectrum in the frequency domain is then

\begin{align}
	T(\omega) & = \left| \frac{\tilde{b}_{out}}{\tilde{b}_{in}}\right|^2 \\
	& = \left| \frac{\sqrt{\gamma_{1} \gamma_{2}}}{-i (\omega - \tilde{\omega}_{C}) + \frac{1}{2} (\gamma_{1} + \gamma_{2}) + \frac{g^2}{-i (\omega - \tilde{\omega}_{X})}} \right|^2
\end{align}
where we introduced the complex eigenfrequencies $\tilde{\omega}_{C} = \omega_{C} - i \kappa$ and $\tilde{\omega}_{X} = \omega_{X} - i \gamma_{X}$ to account for the intrinsic losses of the cavity and exciton, respectively. The result can be simplified to Eq. \ref{trans}.

The minima in the strong coupling regime at zero exciton-cavity detuning ($\omega = \omega_{C} = \omega_{X}$) gives a transmission of 

\begin{equation}
	T = \left ( \frac{\gamma}{\kappa + \gamma} \right )^2 \frac{1}{\left( 1 + C \right)^2}
\end{equation}
where we define the cooperativity as $C \equiv \frac{g^2}{\gamma_{X} (\kappa + \gamma)}$, which effectively quantifies the visibility of the polariton modes. In the absence of an optical transition this reduces to the $T_{max}$ discussed in the main article. There are also maxima in the strong coupling regime at $\omega_{\pm} = \omega_{C} \pm \sqrt{C' g^2 - \gamma_{X}^2}$ with a transmission of

\begin{equation}\label{Tmaxg}
	T = \frac{\gamma^2 C'}{2 g^2 \left[1 - C' \right] + \left[ (\kappa + \gamma)^2 + \gamma_{X}^2 \right] C' + 4 \gamma_{X} \left[ (\kappa + \gamma) + \gamma_{X} \right]}.
\end{equation}
where we define a new constant $C' =  \sqrt{1 + \frac{2}{C} [1 + \gamma_{X} / (\kappa + \gamma)]}$. This is not a particularly illuminating equation other than offering a direct calculation of the transmission maximum. 


The transmission spectrum for the side-coupled cavity can be similarly derived \cite{fan_input-output_2010, rephaeli_few-photon_2012} to be

\begin{equation}
	T(\omega) = \left|1 - \frac{\gamma_{SC}}{-i (\omega - \tilde{\omega}_{C}) + \gamma_{SC} + \frac{g^2}{-i (\omega - \tilde{\omega}_{X})}} \right|^2
\end{equation}
where $\gamma_{SC}$ is the coupling rate between the waveguide and the cavity.

\bibliography{references}
\end{document}